\documentclass[authoryear,final,3p, times]{elsarticle}

\usepackage{amssymb}
\usepackage{algpseudocode}
\usepackage{algorithm}

\journal{TBD}

\let\today\relax
\makeatletter
\def\ps@pprintTitle{%
    \let\@oddhead\@empty
    \let\@evenhead\@empty
    \def\@oddfoot{\footnotesize\itshape
         {} \hfill\today}%
    \let\@evenfoot\@oddfoot
    }
\makeatother

\begin{document}

\begin{frontmatter}


\author[inst1]{Nicolas Cofre\corref{cor1}}
\cortext[cor1]{Corresponding author, e-mail: nicolas.cofre@gatech.edu}

\title{A simulated electronic market with speculative behaviour and bubble formation}



\affiliation[inst1]{ 
           addressline={}, 
           city={Phd Student. University of Gdansk, Faculty of Economics},
           state={Armii Krajowej 119/121, 81-824, Sopot},
           country={Poland}}

\author[inst2]{Magdalena Mosionek-Schweda \corref{cor2}
}

\cortext[cor2]{e-mail: magdalena.mosionek.schweda@uni.lodz.pl}

\affiliation[inst2]{
           addressline={}, 
           city={Professor. University of Lodz, Faculty of Economics and Sociology},
           state={POW 3/5, 90-255, Lodz},
           country={Poland}}

\begin{abstract}
This paper presents an agent based model of an electronic market with two types of trading agents. One type follows a mean reverting strategy and the other, the speculative trader, tracks the maximum realised return over recent trades. The speculators have a distribution of returns concentrated on negative returns, with a small fraction making profits. The market experiences an increased volatility and prices that greatly depart from the fundamental value of the asset. Our research provides synthetic datasets of the order book to study its dynamics under different levels of speculation.
\end{abstract}


\end{frontmatter}


\section{Introduction}


The phenomenon of bubble formation has received extensive attention in the literature. For instance, in the early paper by \citet{Tirolei1985}, it is stated that a prerequisite for the emergence of a bubble is a shared consensus regarding the asset's value. The experimental work of \citet{moinas2013bubble} and \cite{Janssen2019} shows the importance of considering the behavioural traits of the agents. The bubble appears in an asset without fundamental value. They both used a sequential game and the traders decided to buy and sell the asset, but they were not aware of their position in the sequence. \cite{weitzel2020bubbles} experiments with financial professionals and students and shows that bubbles appear for different market compositions. They also concluded that financial professionals act as price stabilizers when mixed with students. In the model proposed by \citet{palfrey2012speculative}, the emergence of a bubble is driven by the heterogeneity in agents' updating strategies after receiving new information. When considering a Bayesian updating rule as the reference, certain agents exhibit a tendency to under-react, while others tend to over-react to the incoming news. In \citet{abreu2003bubbles} there is heterogeneity with respect to the time that arbitrageurs realise that there is a bubble. Bubbles can persist because the rational arbitrageurs players that are aware of the bubble are trying to profit from it for some time and therefore cannot coordinate their selling and burst it as soon as detected.



Our work attempts to contribute to the bubble formation literature simulating a number of agents interacting in a market with explicitly characterized motivations for each trade, incorporating different sources of heterogeneity in the same model. In our paper, the speculators are agents with lottery-like preferences. This type of preference for assets with high maximum returns or lottery-like demand, which is defined in \citet{bali2011maxing} as the demand for low probability/high payoff assets, has been studied in \cite{Lu2022} for stocks and in \cite{Grobys2021} for cryptocurrencies. The common proxy for lottery-like demand is the use of the maximum return observed or MAX factor. We provide a model with speculative agents that generates bubbles and it is useful for controlled experimentation. This simulated market environment could serve as an experimental platform for the investigation of methods for the detection of speculative bubbles. For recent developments in bubble detection methods see the works of \citet{feng2023testing}, \citet{pedersen2020testing}, \citet{astill2023using}, \citet{wang2023testing} and \citet{caravello2023rational}. Another application involves the pricing done by the market maker and how it reacts to the presence of a bubble, testing algorithmic pricing methods as in \citet{klein2021autonomous}. Our agent based model is based on the ABIDES-markets implementation by \cite{byrd2019abides}. For its extension compatible with OpenAI Gym see \cite{amrouni2021abidesgym}. For OpenAI Gym and the current version re-named Gymnasium see \citet{towers_gymnasium_2023}, which allows for the training of agents using reinforcement learning. For other works using ABIDES see \cite{balch2019}, which uses the simulator to assess the market impact of order sizes. \cite{cao2022a}, which develops a synthetic dataset based on the simulations. \cite{10.1145/3531056.3542774} uses it as a guideline for synthetic dataset generation.

An overview of agent-based models and examples of applications in finance can be found in \citet{axtell2022agent}. For other works using agent based models see
\citet{geanakoplos2012getting},
\citet{lebaron2000agent},
\citet{steinbacher2021advances},
\citet{rekik2014agent},
\citet{wang2018agent},
\citet{m2020behavioral},
\citet{katahira2019development},
\citet{el2022agent}
and
\citet{poggio2001agent}.

\section{The model}

The simulations contain three agent types: mean reverting agent, speculative agent and a market maker. They can trade one asset in a limit order book. The agents involved and the asset's value process are described below.

\subsection{Trading agents}

The model involves trading agents of two types, they all start with 10 million in cash. One type of agent has a mean reverting strategy. It has a valuation for the asset and buys if the valuation is above the current mid price and sells otherwise. The implementation for this agent differs from the value agent provided in the source code of \cite{byrd2019abides} in the updating rules. In our implementation, the agent relies on its current belief of the long term fundamental value and in the previous estimates of the short term value, capturing the idea of having different reactions to news. In one extreme the agent will set the belief as the latest observed value while in the other, the agent would never update this belief. The agent applies mean reversion of the short term estimate to the long term value. The agent does not keep estimates of the variance of the estimations nor has access to the variance of the fundamental value observations among agents as in \cite{byrd2019abides}. 

The speculative agent type keeps the history of recent mid prices and will place a buy order if the recently observed maximum log return is high enough. The speculative agent would sell after hitting a take profit or stop loss level. Short selling is not allowed for the speculative agent, but they can use leverage.

The fact that we consider only two types of agents is further motivated by \citet{algorithms2017collusion}, which comments on the problems that algorithmic trading might pose for competition and how tacit collusion might be facilitated with the increased use of machine learning or black box models in trading. In our opinion, the release of public tools for the development of trading strategies\footnote{For example: https://www.composer.trade/learn/best-ai-trading-bot-platforms, which declares to use chatGPT4} makes any coordination less costly, as different agents can get access to the same under-laying code or type of strategy for trading. Our speculative agent serve as a representation of less sophisticated traders, motivated by the findings on the destabilizing effect of retail traders from \citet{baig2023reprint}.




There is heterogeneity regarding the timing of the trading, as each agent has a time between trades exponentially distributed. The mean reverting agents, also present heterogeneity in terms of the mean reversal belief, the reaction to news and about the fundamental value of the asset. Therefore, some agents might choose to ride the bubble for longer than others depending on their beliefs. The beliefs of the value are normally distributed, the mean reversion speed and reaction to news are uniformly distributed. The speculators are heterogeneous with respect to the window for the maximum return calculation. This window is defined by the number of periods that the agents look back searching for the maximum return observed. This number of periods is Poisson distributed. The take profit, stop loss and returns triggering a buy order thresholds are also different for each agent and exponentially distributed. Following \cite{abergel2013mathematical}, the order size or volume for each agent is  drawn from a log-normal distribution.

\begin{algorithm}[H]
\caption{Speculative agent's strategy}\label{alg:cap}
\begin{algorithmic}
\State At inception, generate agent's parameters 
\State $\alpha \sim$ Exp, $\beta \sim$ Exp, $\ r_b \sim$ Exp, $N \sim$ Poisson
\State Generate waiting time to enter the market, $\Delta \sim$ Exp

\While {$t+\Delta< T_{final}$}

\State Generate order size $\sim$ Lognormal

\If{$W_t > \alpha W_{LT} \lor W_t<\beta W_0$}

    \State Send market sell order
\ElsIf{$MAX_N \geq r_b$}
   \State Send market buy order 
   \State update $W_{LT}$
\EndIf

\State Generate waiting time to enter the market, $\Delta \sim$ Exp

\EndWhile
\end{algorithmic}
\end{algorithm}

$W_{LT}$ is the mark to market value of the holdings at the moment of the last buy trade. $W_t$ is the current mark to market value of the holdings and $W_0$ is the starting cash. $T_{final}$ is the final time of the simulation. $MAX_N$ represents the maximum log return over the last $N$ mid prices.


\subsection{Market maker agent}

There is a market maker that generates orders on the ask and bid sides of the order book. As stated in \cite{mcgroarty2019high} the market maker attempts to profit from the spread in exchange of being a liquidity provider. This agent periodically resubmits a set of orders at different price levels (ticks). Considering the previous spreads using an exponentially weighted moving average (EWMA) this agent places orders such that the difference between levels is proportional to the EWMA spread and the volume or size placed at each level is proportional to the traded volume. As mentioned in \cite{balch2019} the market maker model implemented in ABIDES is similar to the one in \cite{chakraborty2011market}, but the ladder is built with a spacing proportional to the spread, instead of unit spacing. In our implementation the market maker skews the prices symmetrically around the mid price based on the current holdings of the asset. As mentioned in \cite{czupryna2022market}, if the market maker's flow is skewed towards buying orders and thus building positive inventory, the market maker would decrease both the bid and ask quotes. Analogously, if the flow is skewed towards selling orders, it would increase both the ask and bid quotes. Price skewing by a market maker is also considered in \cite{bergault2021mathematical} and \cite{Barzykin2021}.







\subsection{The asset fundamental value}
The asset fundamental value follows an OU mean-reverting process which is affected by additional shocks. Those shocks follow a mixture of 2 equally weighted normal distributions, with the same absolute mean, representing positive and negative shocks to the value of the asset. 
The agents can get noisy observations of the value that are normally distributed around the true fundamental value, which creates heterogeneity about the value beliefs. The speculative agents base their trading strategies on the observed returns and not on the fundamental value beliefs.



\section{Results}



\subsection{Market without speculative agents}

In this section we describe the results for the market that is populated with 500 mean reverting agents. As it can be seen in the figure below (random seed 2023), they generate a price path that follows closely the fundamental value of the asset.
\begin{figure}[H]
\caption{Simulation with 500 mean reverting agents}
\includegraphics[width=\textwidth]{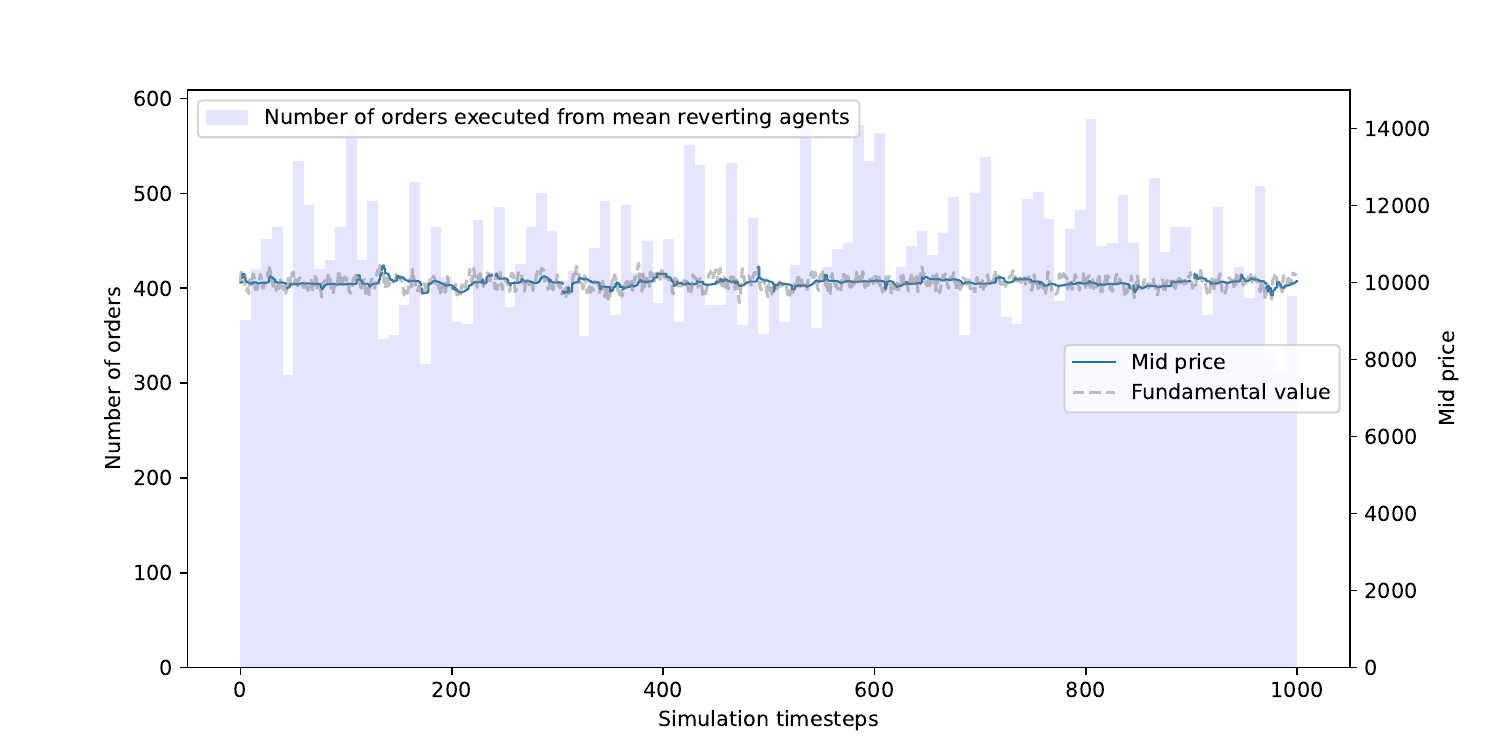}
\end{figure}





The following figure shows the evolution of the portfolio mark to market for each agent. In this scenario all the agents follow a mean reverting strategy.
\begin{figure}[H]
\caption{MtM values for each agent in the scenario with mean reverting agents}
\includegraphics[width=\textwidth]{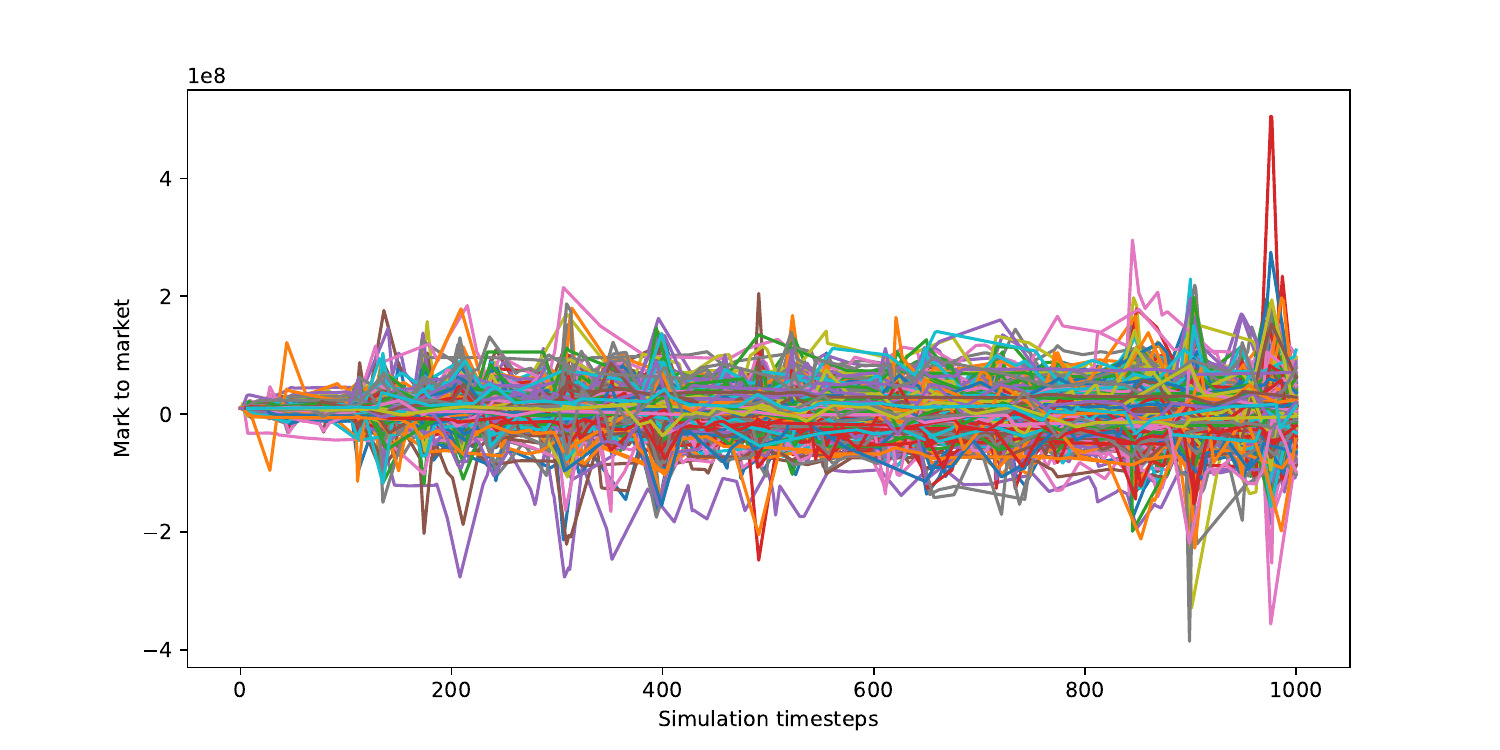}
\end{figure}

The mean reverting agents produce a price that follows closely the true fundamental value of the asset. The heterogeneity among agents allows for different levels of profitability as they have different beliefs regarding the fundamental value dynamics, but no bubble is created in this scenario.


\subsection{Market with speculators}

This sections presents the results for a market that contains the same fundamental value for the asset (same random seed), but this market has been created with 100 mean reverting agents and 400 speculative agents. In this case, the mid price shows bubble behaviour when compared against the fundamental value of the asset. This can be seen in the figure below.
\begin{figure}[H]
\caption{Simulation with 100 mean reverting agents and 400 speculators}
\includegraphics[width=\textwidth]{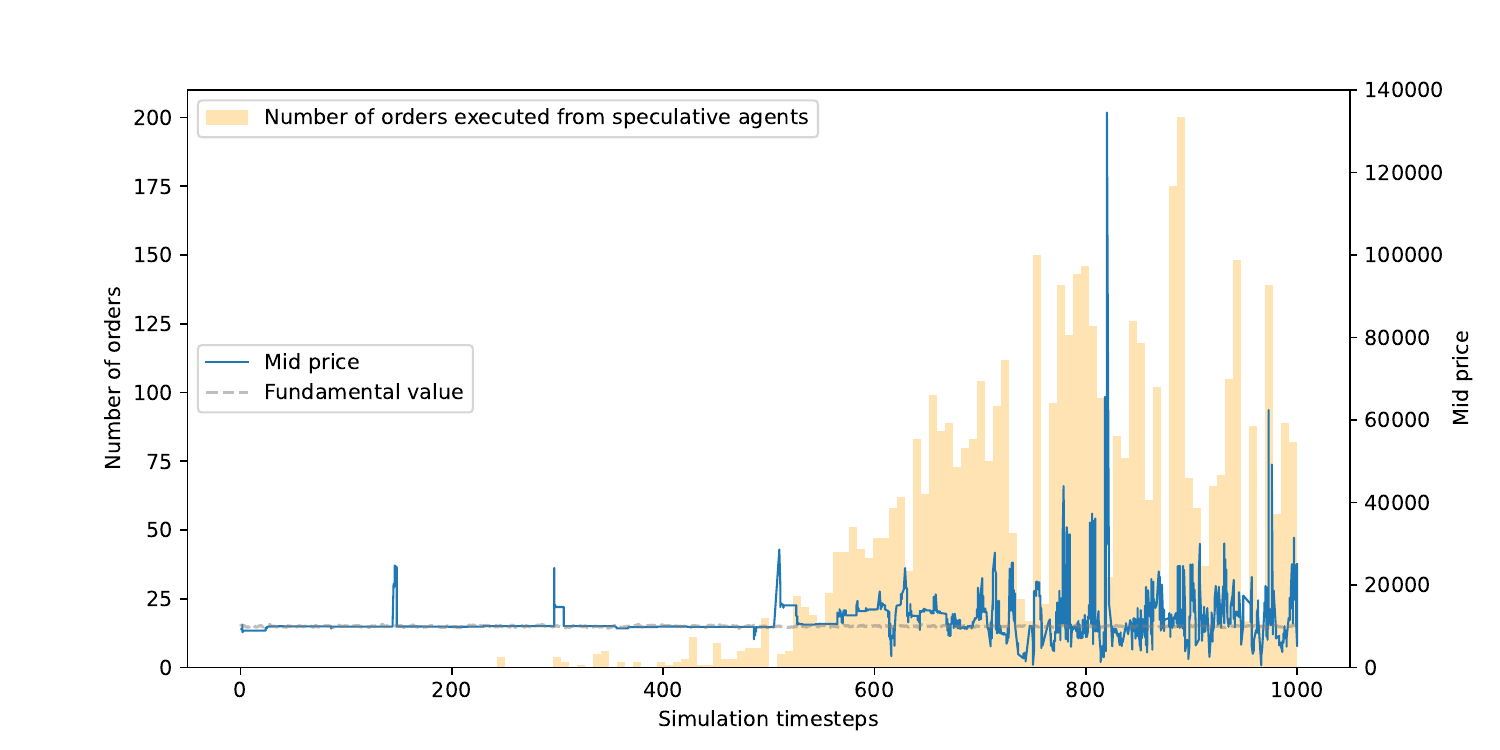}
\end{figure}






During the first half of the simulation, there is almost no trading done by the speculators and there is an increased volatility among the trades executed by the 100 mean reverting agents. Compared to the baseline case of 500 mean reverting agents, we observe short-lived bubbles among the heterogeneous mean reverting agents. But the price quickly returns to levels close to the fundamental value. In the second half of the simulation, the speculative agents start to engage in trading, producing another period of high volatility. This increased volatility promotes the entrance of more speculative agents, which in turns increases the demand for the asset, creating an imbalance in the inventory of the market maker. The market maker's response is to increase both the ask and bid prices of its orders. The bubble during the second part of the simulation is the effect of both the pricing ladder of the market maker reacting to a trade imbalance and the higher demand of the speculative agents.

The onset of the speculative trading occurs several simulated timesteps before the peak of the bubble. An open question for future research is how fast the bubble can be anticipated from the activity in the order book data.

The time series of each agent's mark to market values during the simulation show the under-performance of the speculators. For the speculative agents these are skewed towards negative values as it can be seen in the next figure.
\begin{figure}[H]
\caption{MtM for the mean reverting and speculative agents}
\includegraphics[width=\textwidth]{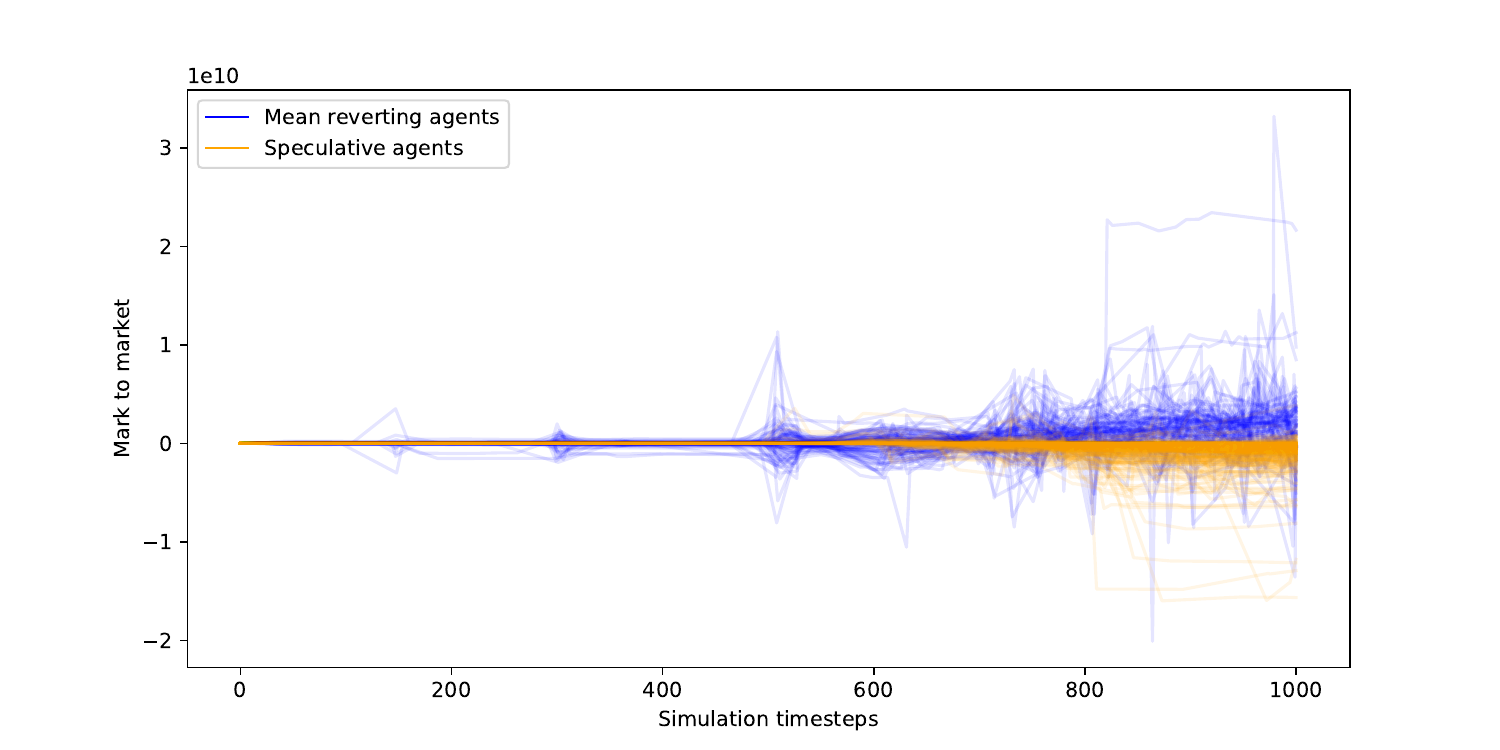}
\end{figure}




\section{Conclusions}


This paper provides insights from an agent-based point of view on the complex dynamics of the market microstructure that drives bubble formation in speculative markets.
Our work provides a source for synthetic datasets generation, which is useful for testing trading or risk models, pricing models from the market maker perspective and also for the research of bubble detection methodologies.

With the easier access to algorithmic trading platforms by retail traders, it is important to understand the effect of limited rationality in the markets. As seen in the simulations of this paper, with the increased presence of less sophisticated traders, short-sighted profit looking strategies can create a self-confirming loop in which a shock causing a positive return might incentivize more demand solely based on the shock and not on the fundamentals, further exacerbating the effect of the shock.



Under this controlled environment further research would explore the effects of more agent types and different variables on the bubble formation, like the degree of heterogeneity of beliefs among the traders, the speed of trading, liquidity constraints, leverage constraints and the pricing of the market maker. 

In our simulated market, the speculative activity builds up several timesteps before the peak of the bubble. Using the simulated order book data, further research would focus on its early detection and prevention to help increase the stability of electronic markets.  



\section{Data availability}
Data will be made available on request.

\section{Declaration of competing interest}
We declare that we have no relevant or material financial interests that relate to the research described in this paper.

 \bibliographystyle{elsarticle-num-names} 
 \bibliography{paper_draft}





\newpage
\appendix
\section{Simulations with other random seeds}

\begin{figure}[H]
\caption{Mid prices for different random seeds. Case with 100 mean reverting agents and 400 speculators. 20 simulations with seeds $\in \{2023,...,2042\}$}
\includegraphics[width=\textwidth]{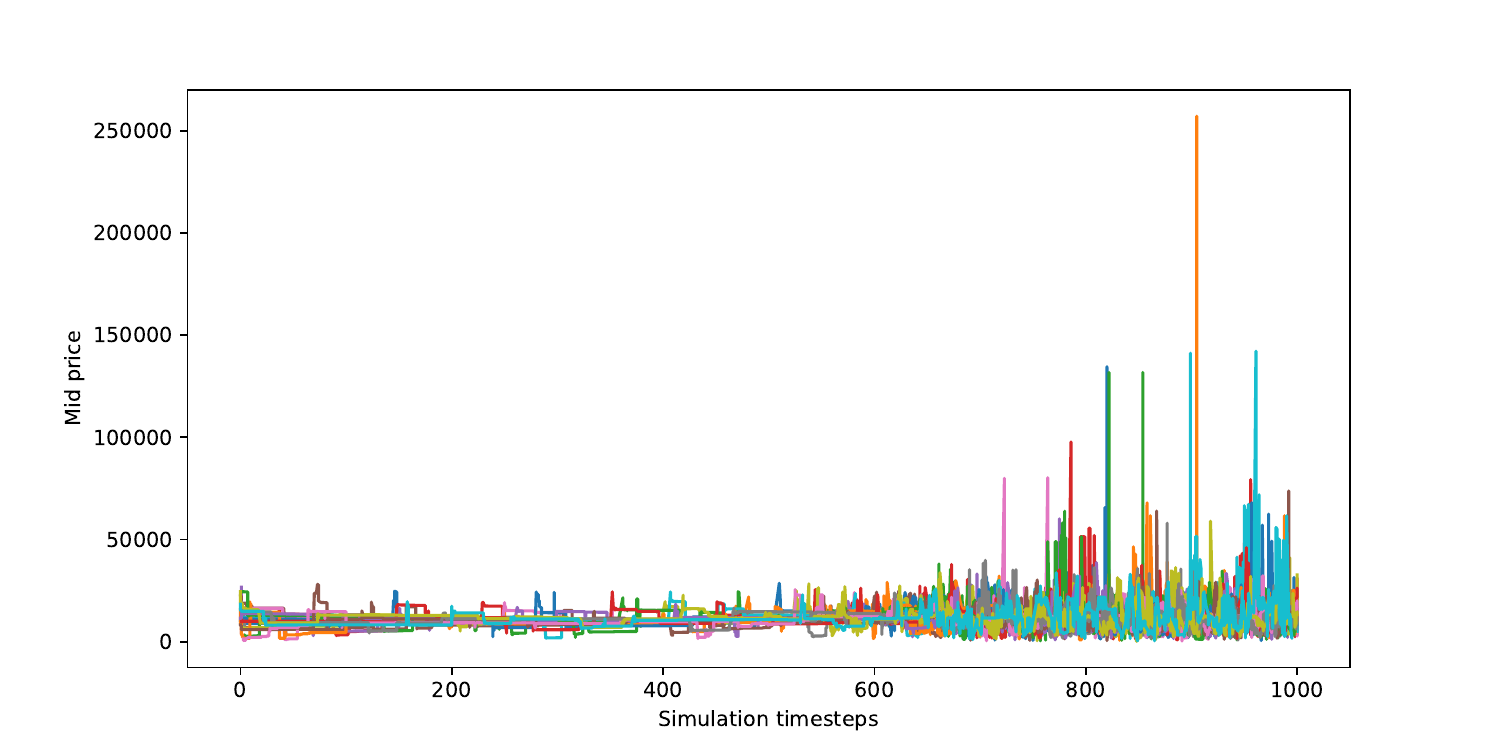}
\end{figure}

\section{Simulations with other compositions of agents}


\begin{figure}[H]
\caption{Mid price and fundamental value in a simulation with 200 mean reverting agents and 300 speculative agents }
\includegraphics[width=\textwidth]{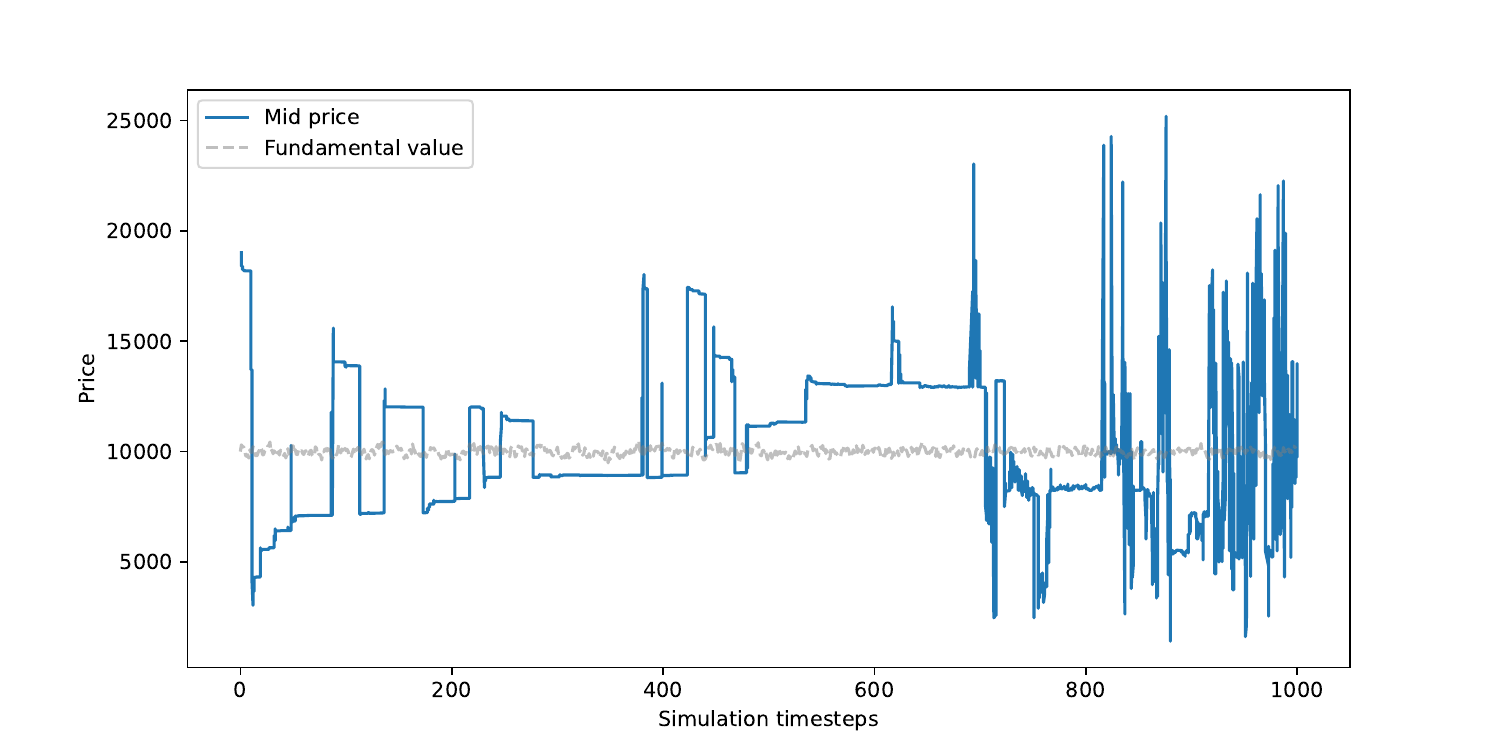}
\end{figure}

\begin{figure}[H]
\caption{Mid price and fundamental value in a simulation with 250 mean reverting agents and 250 speculative agents }
\includegraphics[width=\textwidth]{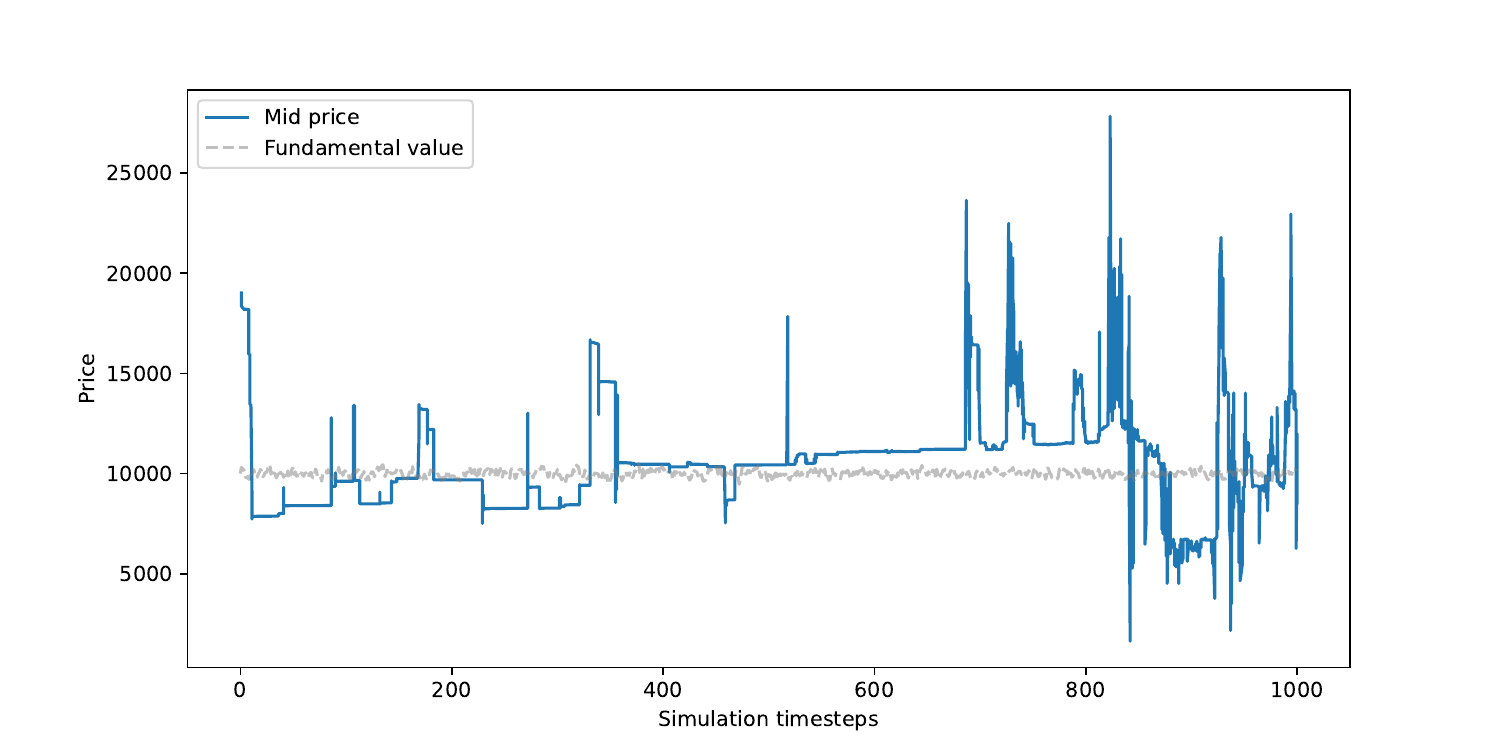}
\end{figure}

\begin{figure}[H]
\caption{Mid price and fundamental value in a simulation with 300 mean reverting agents and 200 speculative agents}
\includegraphics[width=\textwidth]{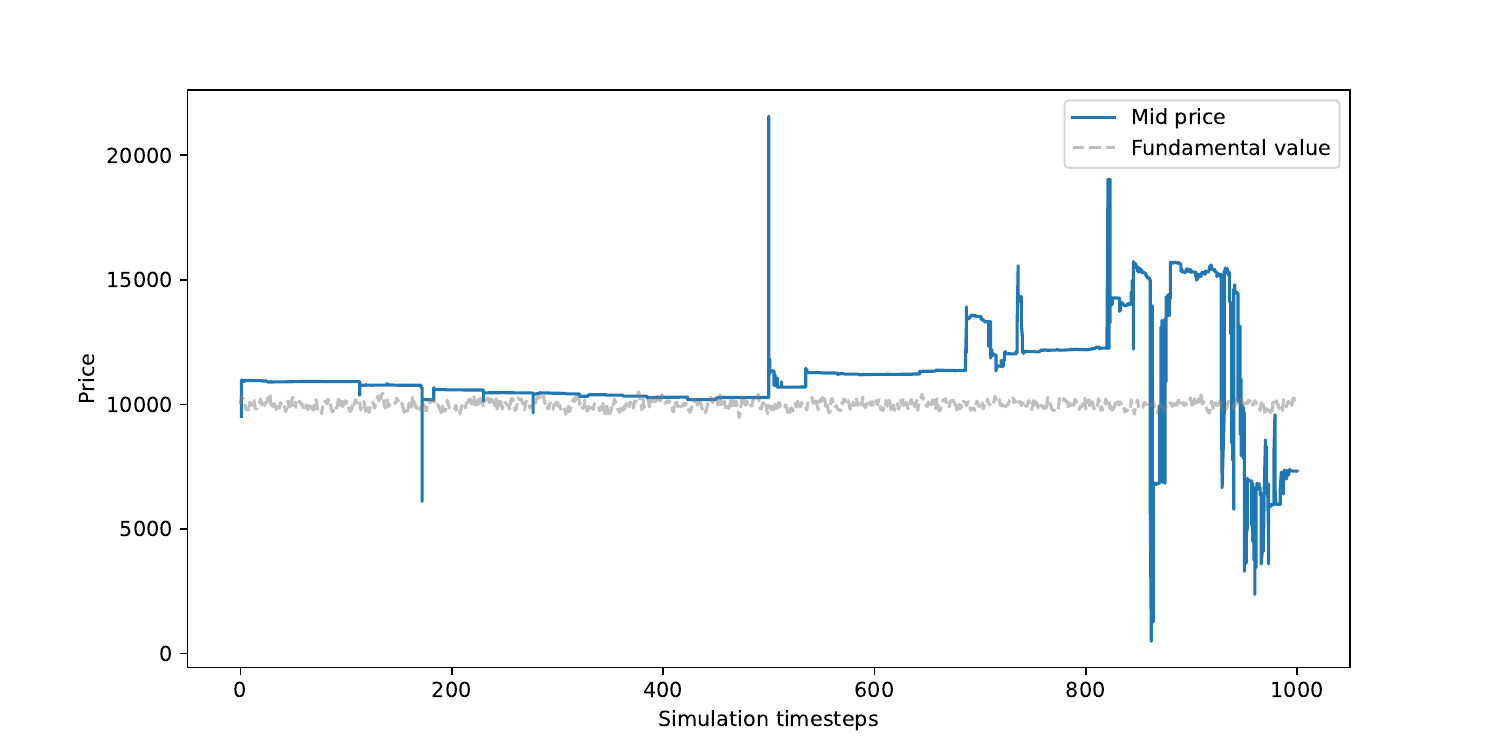}
\end{figure}

\begin{figure}[H]
\caption{Mid price and fundamental value in a simulation with 400 mean reverting agents and 100 speculative agents }
\includegraphics[width=\textwidth]{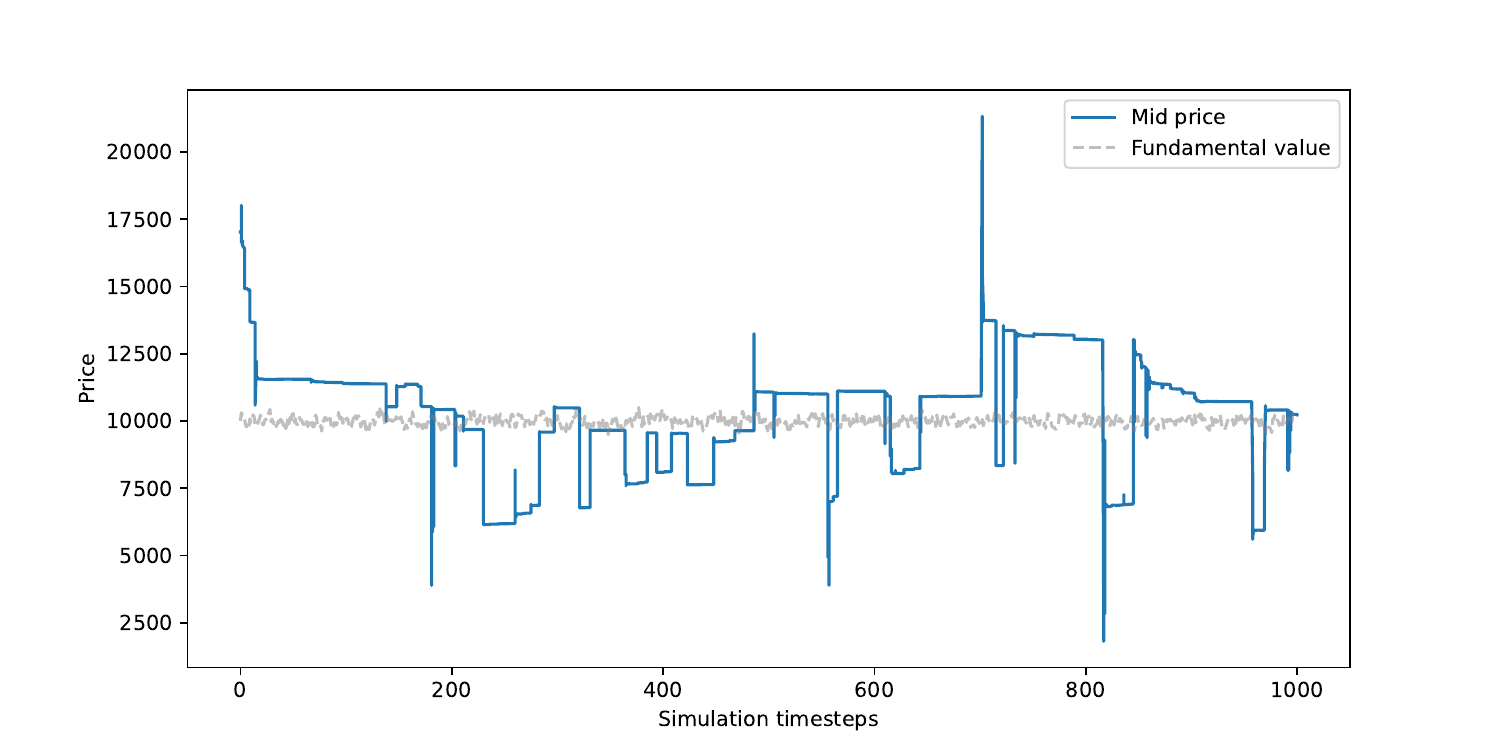}
\end{figure}







\section{Final distribution of mark to market portfolio value}

\begin{figure}[H]
\caption{Final distribution of MtM with 500 mean reverting agents}
\includegraphics[width=\textwidth]{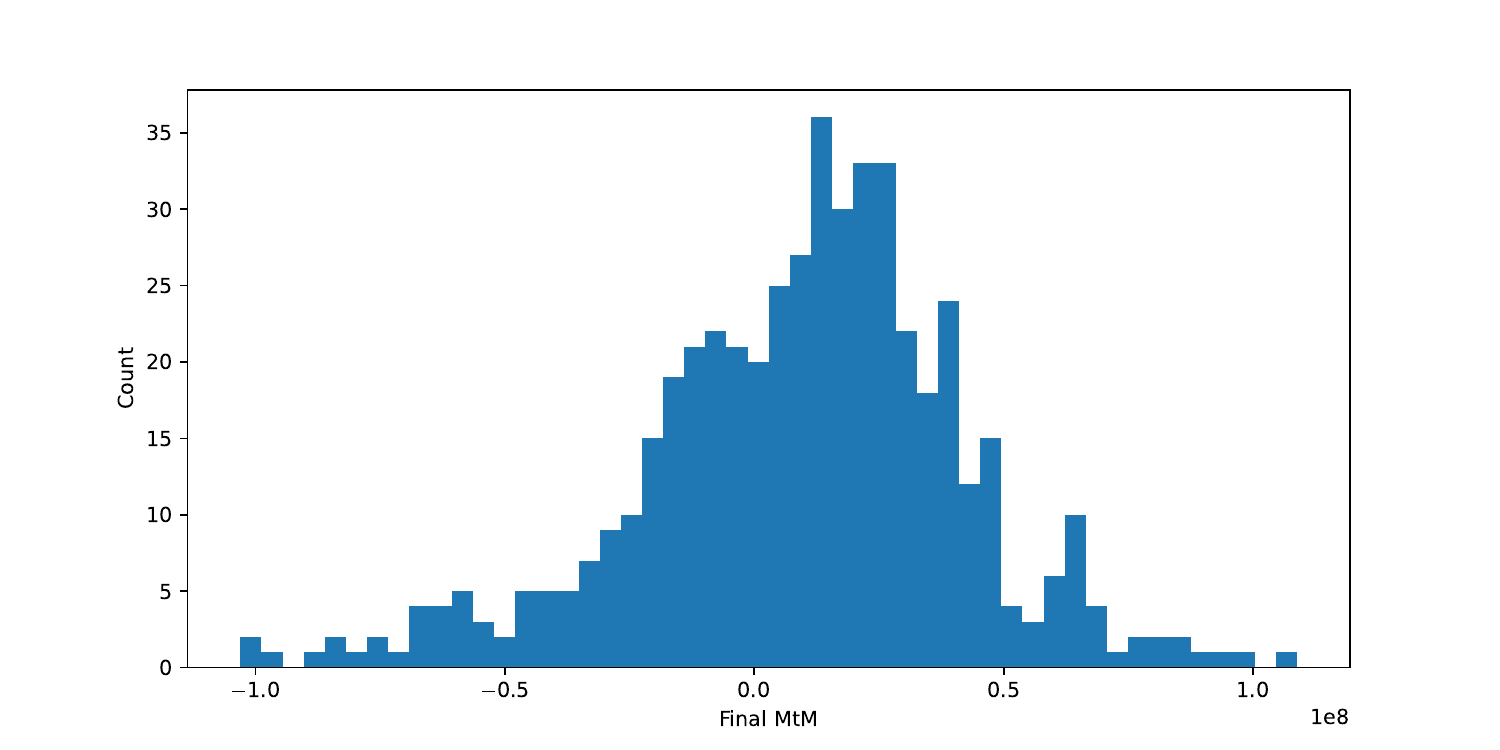}
\end{figure}

\begin{figure}[H]
\caption{Final distribution of MtM with 100 reverting agents and 400 speculative agents}
\includegraphics[width=\textwidth]{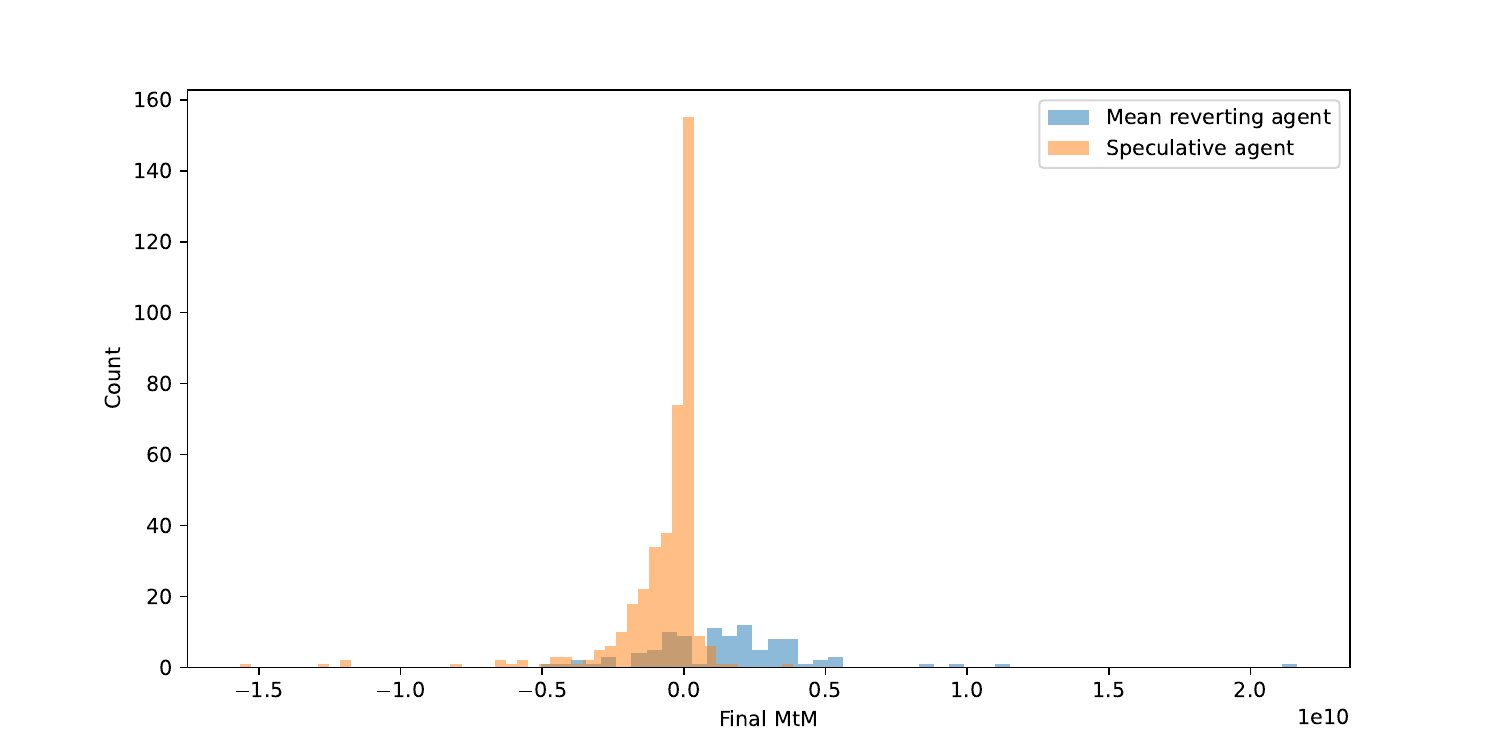}
\end{figure}

\end{document}